\documentclass[11pt,a4paper]{article}

\pdfoutput=1

\usepackage{comment}
\usepackage{jcappub}

\usepackage{physics}
\usepackage{subfig}
\usepackage{float}
\usepackage{xcolor}
\usepackage{appendix}

\newcommand{\be}{\begin{equation}}
\newcommand{\ee}{\end{equation}}
\newcommand{\bea}{\begin{eqnarray}}

\newcommand{\eea}{\end{eqnarray}}

\newcommand{\MP}{M_\text{P}}

\newcommand{\FTtwo}{F_{>2}(R_X)}

\title{\centering Natural inflation in Palatini $F(R,X)$}

\author[a, b]{N. Bostan,}
\author[c]{R. H. Dejrah,}
\author[d,e]{C. Dioguardi,}
\author[e]{and A. Racioppi}

\affiliation[a]{Department of Physics and Astronomy, University of Iowa, 52242 Iowa City, IA, USA}
\affiliation[b]{Proton Accelerator Facility, Turkish Energy Nuclear and Mineral Research Agency, Nuclear Energy Research Institute, 06980, Ankara, Türkiye}
\affiliation[c]{Department of Physics, Faculty of Sciences, Ankara University,  06100, Ankara, Türkiye}
\affiliation[d]{Tallinn University of Technology, Akadeemia tee 23, 12618 Tallinn, Estonia}
\affiliation[e]{National Institute of Chemical Physics and Biophysics, R\"avala 10, 10143 Tallinn, Estonia}

\emailAdd{nilay.bostan@tenmak.gov.tr}
\emailAdd{rafid.dejrah@gmail.com}
\emailAdd{christian.dioguardi@kbfi.ee}
\emailAdd{antonio.racioppi@kbfi.ee}

\abstract{In the context of Palatini gravity, $F(R+X)$ models, with X the inflaton kinetic term, are characterized by the appealing property of generating asymptotically flat inflaton potentials, exactly like the more commonly studied Palatini $F(R)$ models, but without the complication of non-canonical inflaton kinetic terms in the Einstein frame. In this paper, we study the case of a Jordan frame potential which is positive and bounded, specifically, natural inflation. We compute the CMB observables and show that for a wide class of $F(R + X)$ theories, including the quadratic one, natural inflation is still viable.}
\begin{document}
\maketitle
\section{Introduction}

As observations of cosmic microwave background radiation (CMB) demonstrate, our universe is flat and homogeneous at large scales. By introducing an accelerated period of expansion in the early universe before the hot big bang, it is possible to generate the observed flatness and homogeneity without fine-tuning of initial conditions \cite{Starobinsky:1980te,Guth:1980zm,Linde:1981mu,Albrecht:1982wi}. This period of accelerated expansions is called inflation. In the simplest version, inflation is driven by the quasi-constant energy density of the inflaton, a scalar particle embedded in General Relativity (GR). However, most of the simplest models are strongly disfavored by the current CMB measurements \cite{Planck2018:inflation,BICEP:2021xfz}. Among such models there is natural inflation, first introduced in \cite{Freese:1990rb}, which naturally solves the fine-tuning problem of the inflaton potential. There have been several attempts in order to restore its viability (e.g. \cite{Kim:2004rp,Visinelli:2011jy,Achucarro:2015rfa,Ferreira:2018nav,Antoniadis:2018yfq,Salvio:2019wcp,Simeon:2020lkd,McDonough:2020gmn,Salvio:2021lka,Salvio:2022mld,Bostan:2022swq,Salvio:2023cry,Mukuno:2024yoa,Racioppi:2024zva,Lorenzoni:2024krn,Michelotti:2024bbc} and refs. therein). Notably several of them rely on non-minimal formulations of gravity, as they give more freedom in formulating the theory and exploring the available parameter space (e.g. \cite{Jarv:2016sow,Jarv:2020qqm} and refs. therein). Among them, non-minimally coupled to gravity models in the Palatini formulation  have received a lot of attention recently, e.g. \cite{Tamanini:2010uq,Bauer:2010jg,Rasanen:2017ivk,Tenkanen:2017jih,Racioppi:2017spw,Markkanen:2017tun,Jarv:2017azx,Racioppi:2018zoy,Kannike:2018zwn,Enckell:2018kkc,Enckell:2018hmo,Rasanen:2018ihz,Bostan:2019uvv,Bostan:2019wsd,Carrilho:2018ffi,Almeida:2018oid,Takahashi:2018brt,Tenkanen:2019jiq,Tenkanen:2019xzn,Tenkanen:2019wsd,Kozak:2018vlp,Antoniadis:2018ywb,Gialamas:2019nly,Racioppi:2019jsp,Rubio:2019ypq,Edery:2019txq,Lloyd-Stubbs:2020pvx,Das:2020kff,McDonald:2020lpz,Shaposhnikov:2020fdv,Enckell:2020lvn,Gialamas:2020snr,Karam:2020rpa,Gialamas:2020vto,Karam:2021wzz,Karam:2021sno,Gialamas:2021enw,Annala:2021zdt,Racioppi:2021ynx,Cheong:2021kyc,Mikura:2021clt,Ito:2021ssc,Racioppi:2021jai,AlHallak:2021hwb,AlHallak:2022gbv,Gialamas:2022gxv,Dimopoulos:2022rdp,Dimopoulos:2022tvn,Gialamas:2022xtt,Racioppi:2024pno,
Gialamas:2024uar,Bostan:2024ugi,Dioguardi:2021fmr,TerenteDiaz:2023kgc,Kannike:2023kzt,Racioppi:2022qxq}.
In the metric formulation, the only dynamical degree of freedom is the metric, and the affine connection is assumed to be the Levi-Civita one. On the other hand, in the case of Palatini gravity, the affine connection and the metric tensor are considered a priori independent, and the relation between them is set by the corresponding equations of motion (EOMs). The two formulations have been proved to be equivalent in the case of minimally coupled gravity. However, in  presence of non-minimal theories of gravity, the two formulations are not equivalent anymore and yield different phenomenological predictions \cite{Koivisto:2005yc,Bauer:2008zj}.

In this study, we are interested in a particular class of non-minimal Palatini models: the $F(R,X)$ models, with $X$ the inflaton kinetic term. This kind of theories has been introduced in \cite{Dioguardi:2022oqu} as way to heal the Palatini\footnote{Analoguosly, $F(R)_{\leq 2}$ will stand for the theories containing terms diverging not faster than $R^2$.} $F(R)_{>2}$ (that is those containing terms diverging faster than $R^2$) class of models, which has shown to have dynamical issues with the evolution of the scalar field outside slow-roll due to the presence of higher-order kinetic terms in the Einstein frame. The remarkable feature of the Palatini $F(R)_{>2}$  theories is the positivity of Einstein frame inflaton potential, regardless of the initial sign of the Jordan frame one \cite{Dioguardi:2021fmr,Dioguardi:2022oqu}.  Such a property is shared by the Palatini $F(R,X)_{>2}$ models, with the additional bonus of simplified inflaton kinetic terms in the Einstein frame \cite{Dioguardi:2022oqu,Dioguardi:2023}. Therefore, n this context, inflation  has been extensively studied in \cite{Dioguardi:2022oqu,Dioguardi:2023,Dioguardi:2025vci} for both positive and negative unbounded inflaton potentials in the Jordan frame, as an extended generalization of the $F(R)$ models already explored in \cite{Dioguardi:2021fmr}. Moreover $F(R, X)$ models seems to provide a natural framework for quintessential inflation realizations \cite{Dioguardi:2022oqu,Dioguardi:2023,Dimopoulos:2025fuq}.

The main purpose of this manuscript is to study the remaining complementary case, that is a positive bounded potential in the Jordan frame embedded in $F(R,X)$ Palatini gravity. In particular, we choose the case of natural inflation. The paper is organized as follows. In section \ref{sec:Palatini}, we introduce the class of Palatini $F(R, X)$ models, and describe their fundamental features. In section \ref{sec:n<2}, we compute the CMB observables for $F(R,X)_{\leq 2}$ while in section \ref{sec:n>2} we do the same but for the $F(R,X)_{>2}$ case. In section \ref{sec:conclusion}, we summarize the results and draw our conclusion. Finally, some additional details about slow-roll computation are given in Appendix \ref{appendix:SR}.
\section{Palatini $F(R,X)$ theories}\label{sec:Palatini}
In the following, we assume Planck units, where $\MP$ is set equal to unity, and a space-like metric signature. 
The starting point for the $F(R,X)$ action has the form:
\be \label{eq:actionFR}
  S_J = \int \dd^4 x \sqrt{-g_J} \left[\frac{1}{2} F(R(\Gamma),X) - V(\phi) \right] \, ,
\ee
where $X$ is the inflaton kinetic term given by $X\equiv- g_J^{\mu\nu} \partial_\mu \phi \partial_\nu \phi$,
where the use of the Palatini formulation of gravity (and therefore an independent affine connection $\Gamma^\rho_{\mu\nu}$) is emphasized in the notation ``$R(\Gamma)$" for the curvature scalar. In this work, we follow the simplest case for which $F(R,X) = F(R_X)$ with $R_X\equiv R+X$.
This is the simplest choice that allows to solve issue of the higher order kinetic terms for the inflaton in Einstein frame. Such a setup has been extensively studied in \cite{Dioguardi:2023,Dioguardi:2022oqu,Dimopoulos:2025fuq,Dioguardi:2025vci}, therefore we summarize in the following only the most important details.
By introducing an auxiliary field $\zeta$, we can rewrite the action as:
\bea \label{eq:action:zeta:J}
  S_J &=& \int \dd^4 x \sqrt{-g}_J \left[\frac{1}{2} \left[F(\zeta)+F'(\zeta) \left(R_X-\zeta \right) \right] - V(\phi) \right] \, ,
\eea
where we used $F'(\zeta)\equiv\partial F/\partial \zeta$. By performing a Weyl transformation we can move to the Einstein frame,
\be \label{eq:Weyl}
  g_{E\, \mu \nu} = F'(\zeta)  \ g_{J \, \mu \nu} \, ,  
\ee
which gives
\be \label{eq:action:zeta:E}
  S_E = \int \dd^4 x \sqrt{-g_E} \left[ \frac{1}{2} R_E - \frac{1}{2} g_E^{\mu\nu} \partial_\mu \phi \partial_\nu \phi - U(\phi,\zeta) \right] \, ,
\ee
where we notice that the canonically normalized scalar field $\phi$ is the same that appeared in the Jordan frame inside the argument of the $F$ function. The scalar field potential takes the form:
\be \label{eq:Uchizeta}
 U(\phi,\zeta) = \frac{V(\phi)}{F'(\zeta)^2} - \frac{F(\zeta)}{2 F'(\zeta)^2} + \frac{\zeta}{2 F'(\zeta)} \, .
\ee
 By varying \eqref{eq:action:zeta:E} with respect to $\zeta$, we get its EOM in the Einstein frame,
\be
G(\zeta) = V(\phi) \, , \label{eq:EoMzetafull}
\ee 
with
\be
G(\zeta) \equiv \frac{1}{4}[2F(\zeta)-\zeta F'(\zeta)].
\ee
Applying the computational strategy introduced in \cite{Dioguardi:2021fmr} (i.e. substituting $V(\phi)$ with $G(\zeta)$ in \eqref{eq:Uchizeta}, one can show that the Einstein frame inflaton potential can be formally written as a function of $\zeta$ only\footnote{We remind that eq. \eqref{eq:U} was originally found in \cite{Dioguardi:2021fmr} in the Palatini $F(R)$ framework. However in that case, eq. \eqref{eq:EoMzetafull} is an approximated result valid only under slow-roll. Now, in the Palatini $F(R+X)$ framework, eq. \eqref{eq:EoMzetafull} is instead exact.}:
\be \label{eq:U}
 U(\zeta) =\frac{1}{4} \frac{\zeta}{F'(\zeta)} \, .
\ee
In general, \eqref{eq:EoMzetafull} can't be solved explicitly for $\zeta$, however, we can still compute the CMB observables by using the formalism developed in \cite{Dioguardi:2021fmr} i.e. using $\zeta$ as the computational variable. 
Consistency of the theory requires $F'(\zeta)>0$. If this constraint is satisfied and $\zeta$ is positive, then $U$ is positive definite for any $V(\phi)$, positive or negative, that satisfies eq. \eqref{eq:EoMzetafull}. The case of a positive $V(\phi)$ and unbounded from above and of a negative $V(\phi$) unbounded from below have been already studied in \cite{Dioguardi:2023,Dioguardi:2022oqu}. This time we study the complementary case in which $V(\phi)$ is positive and bounded from above. Specifically, we study the case of a Jordan frame potential given by natural inflation:
\be
 V(\phi) = \Lambda^4 \left[1+\cos{\left(\frac{\phi}{M}\right)}\right] \label{eq:V}.
\ee
In the following, we consider polynomial $F$ of the form
\be
F(\zeta) = \zeta +\alpha \zeta^{n} \, , \label{eq:F}
\ee
and study separately $n\leq 2$ and $n>2$ because of the different properties that $G(\zeta)$ will exhibit when studying the two distinct cases.
\section{Natural inflation for $n\leq 2$}\label{sec:n<2}
The function $G(\zeta)$, obtained from a $F(\zeta)$ given by eq. \eqref{eq:F}, reads
\be\label{eq:G_n}
G(\zeta)=\frac{1}{4}(\zeta + \alpha (2-n)\zeta^n).
\ee
It is then clear that the function is positive and monotonically increasing for any $\zeta>0$ and $1<n\leq 2$. This ensures that the mapping between $G(\zeta)$ and $V(\phi)$ is always possible and, in particular, for $n=2$ it allows to solve explicitly \eqref{eq:EoMzetafull}, giving  $\zeta = 4V(\phi)$. 
We also stress that the passage from $n \neq 2$ to $n=2$ is discontinuous because of the appearance (disappearance) of the $(2-n) \zeta^n$ contribution when $n \neq 2$\ $(n=2)$.
We will start from the latter, given the simplicity of the computations, and then study and compute the CMB observables for $1<n<2$ (in particular $n=\frac{3}{2},\frac{7}{4}$ in order to understand the behavior as $n\rightarrow 2$).

For the $n=2$ case, the Einstein frame potential can be explicitly written as:
\be
U(\phi) = \frac{\Lambda^4\qty(1+\cos\frac{\phi}{M})}{1+8\alpha\Lambda^4\qty(1+\cos\frac{\phi}{M}) }. \label{eq:U:2}
\ee
A potential with a similar form has been already studied in \cite{Antoniadis:2018yfq}, in the context of a quadratic Palatini $F(R)$. However, in ref.\cite{Antoniadis:2018yfq} the scalar field also needs to undergo a canonical normalization, while in our framework $\phi$ is already canonical normalized. Therefore, the inflationary results will differ from the ones of~\cite{Antoniadis:2018yfq}.

\begin{figure}[t]%
    \centering
    {\includegraphics[width=0.7\textwidth]{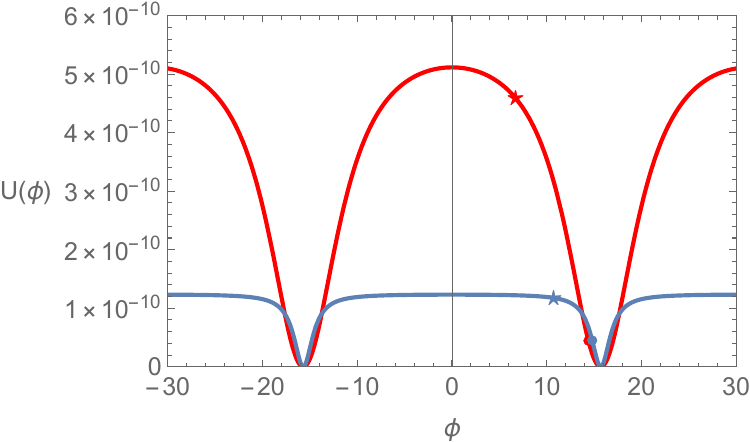}}%
    \caption{Einstein frame potential  for the quadratic $F(R_X) = R_X +\alpha R_X^2$ with $M=5$, and $\alpha=2\cdot 10^8$ (red), $\alpha= 10^9$ (blue), $\Lambda \sim 5\cdot 10^{-3}$ is fixed by imposing the constraint $A_s \simeq 2.1\cdot 10^{-9}$. We also show for reference the field values $\phi_N$ (star) and $\phi_{end}$ (dot). As $\alpha$ increases, the width of the plateaus becomes much larger than the width of the minima of the potential where the field will perform reheating.} 
\label{fig:U_potential_n_2}
\end{figure}

We show an indicative plot of the potential \eqref{eq:U:2} in Fig. \ref{fig:U_potential_n_2} for the choice $M=5$, and $\alpha=2\cdot 10^8$(red), $\alpha=2\cdot 10^9$(blue). $\Lambda \sim 5\cdot 10^{-3}$ is fixed by imposing the constraint $A_s \simeq 2.1\cdot 10^{-9}$. We also show for reference the field values $\phi_N$ and $\phi_{end}$, respectively corresponding to the ``beginning'' and ``end of inflation''. As $\alpha$ increases, the height of the plateaus decreases, and their width becomes much larger than the width of the minima of the potential where the field will perform reheating. The potential is periodic with period $2 \pi M$.
We now proceed with the evaluation of the slow-roll parameters\footnote{For the non-expert reader, a brief summary of the slow-roll formalism is given in Appendix \ref{appendix:SR}.} for the case $n=2$:
\bea
&\epsilon_U(\phi_N)& = \frac{\tan ^2\qty(\frac{\phi_N}{2 M})}{2 M^2 \qty(8 \alpha  \Lambda ^4+8 \alpha  \Lambda ^4 \cos(\frac{\phi_N}{M})+1)^2}, \\
&\eta_U(\phi_N)& = \frac{4 \alpha  \Lambda ^4 \left(\cos \left(\frac{2 \phi_N}{M}\right)-3\right)-\left(8 \alpha  \Lambda ^4+1\right) \cos \left(\frac{\phi_N}{M}\right)}{M^2 \left(\cos \left(\frac{\phi_N}{M}\right)+1\right) \left(8 \alpha  \Lambda ^4+8 \alpha  \Lambda ^4 \cos \left(\frac{\phi_N}{M}\right)+1\right)^2}, \label{eq:eta}
\eea
and the number of efolds reads:
\be
N_e =-2 M^2 \left[4 \alpha  \Lambda ^4 \cos \left(\frac{\phi }{M}\right)+\left(16 \alpha  \Lambda ^4+1\right) \ln \left(\sin
   \left(\frac{\phi }{2 M}\right)\right)\right]_{\phi_{end}}^{\phi_N}
\ee
The corresponding CMB observables are given by:
\begin{align}
r  &= 16 \epsilon_U(\phi_N),\\
n_s  &= 1+2\eta_U(\phi_N)- 6\epsilon_U(\phi_N),\\
A_s  &= \frac{U(\phi_N)}{24\pi^2 \epsilon_U(\phi_N)}.
\end{align}
It can be proven numerically that by increasing $\alpha$ both ${\phi_{end}}$ and ${\phi_N}$ approach $M \pi$ (see also Fig. \ref{fig:U_potential_n_2}). Therefore, by making an expansion at the leading order around $\phi= M \pi$ we can obtain the following approximated results valid in the strong coupling regime:
\begin{align}
r  & \simeq \frac{M}{\sqrt{2} \sqrt{\alpha } \Lambda ^2 N_e^{3/2}} \label{eq:r:RX2:limit} \, ,\\
n_s  & \simeq 1-\frac{3}{2 N_e} + \frac{M}{2 \sqrt{2} \sqrt{\alpha } \Lambda ^2 N_e^{3/2}} \label{eq:ns:RX2:limit} \, , \\
A_s  & \simeq \frac{\Lambda ^2 N_e^{3/2}}{6 \sqrt{2} \pi ^2 \sqrt{\alpha } M} \label{eq:As:RX2:limit}  \, .
\end{align}
For the $n<2$ models, we cannot solve \eqref{eq:EoMzetafull} explicitly hence we have to use the formalism derived in \cite{Dioguardi:2021fmr}, where the auxiliary field $\zeta$ is used as the computational variable.
At large $\alpha$, we get the following expressions for the CMB observables (see Appendix \eqref{eq:epsilon_full}-\eqref{eq:As_full} for the full expressions of the slow-roll parameters and the inflationary observables):

\begin{align}
r(\zeta_N) &\approx \frac{8 (n-2)^2}{M^2 n^2 \left(e^{\frac{(2-n) N_e}{n M^2}}-1\right)} ,\label{eq:r:n:less:2}\\
n_s(\zeta_N) &\approx 1+\frac{(2-n)}{M^2 n^2} \left(\frac{2}{e^{\frac{(n-2) N_e}{M^2 n}}-1}+2-n\right), \label{eq:ns:n:less:2}\\
A_s(\zeta_N) &\approx -\frac{M^2 2^{\frac{6}{n}-7} n^2 e^{-\frac{(n-2) N_e}{M^2 n}}}{3 \pi ^2 \Lambda ^4 (n-2)} \left(\frac{\Lambda ^4 \left(e^{\frac{(n-2) N_e}{M^2
   n}}-1\right)}{\alpha  (n-2)}\right)^{2/n} \, .\label{eq:As:r:n:less:2}
\end{align}
\begin{figure}[!t]%
    \centering
    \subfloat[]{\includegraphics[width=0.45\textwidth]{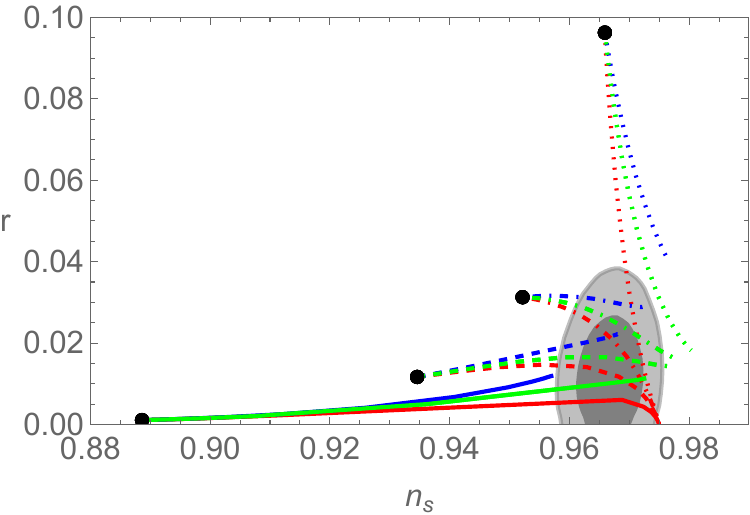}}%
    \qquad
    \subfloat[]{\includegraphics[width=0.455\textwidth]{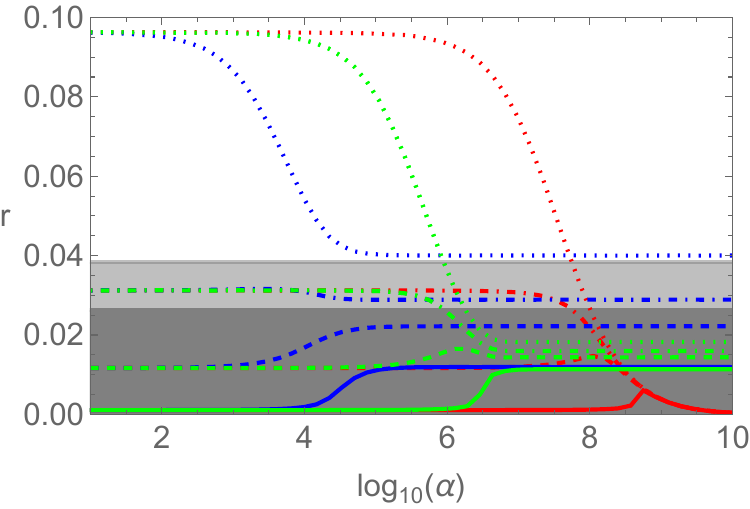}}%
    
    \subfloat[]{\includegraphics[width=0.45\textwidth]{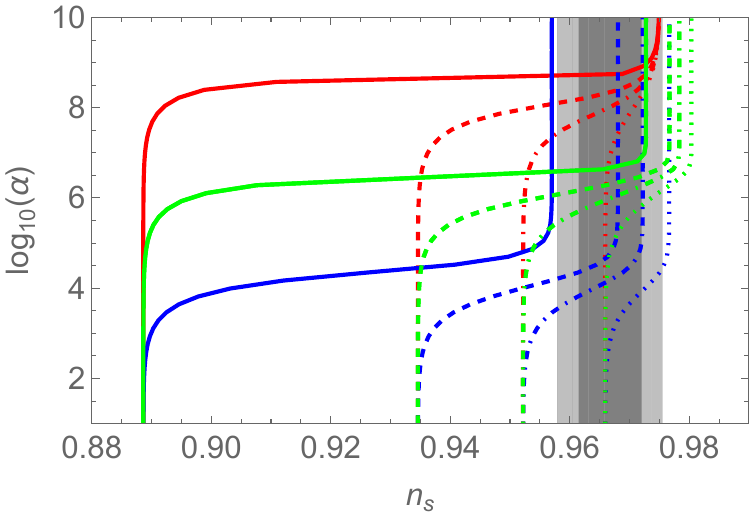}}%
    \qquad
    \subfloat[]{\includegraphics[width=0.475\textwidth]{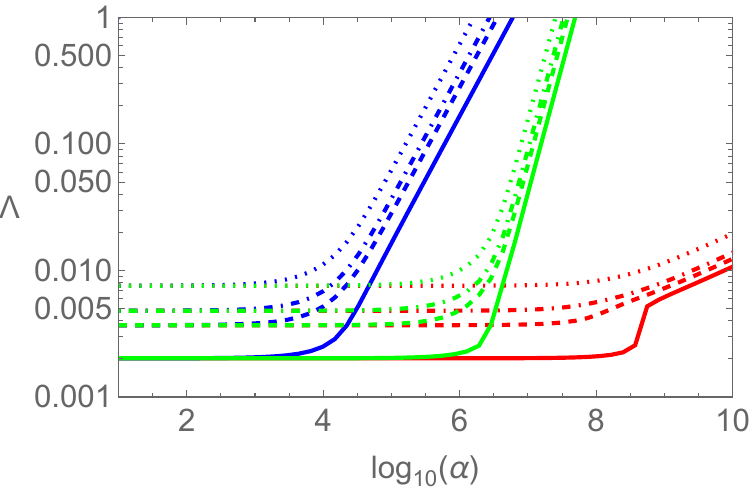}}%

   \caption{(a) $r$ vs. $n_s$, (b)  $r$ vs.  $\log_{10}(\alpha)$, (c) $n_s$ vs. $\log_{10}(\alpha)$, and (d) $\Lambda$ vs.  $\log_{10}(\alpha)$ for the model presented in section \ref{sec:n<2}, with $n=3/2$ (blue)  $n = 7/4$ (green) and $n = 2$ (red) with $M = 3$ (thick), $M = 4$ (dashed), $M = 5$ (dash-dotted), and $M = 10$ (dotted). The black dots show the predictions of the original natural inflation model. All observables are computed at $N_e = 60$. The amplitude of the power spectrum $A_s$ is fixed to its observed value. The gray regions indicate the 95\% (dark-gray) and 68\% (light-gray) confidence levels (CL), respectively, based on the latest combination of Planck, BICEP/Keck, and BAO data \cite{BICEP:2021xfz}.}

    \label{fig:n_less_2}
\end{figure}
The discontinuity from the $n<2$ to the $n=2$ case can be appreciated in eqs. \eqref{eq:r:n:less:2}-\eqref{eq:As:r:n:less:2}, where $r$, $n_s$ and $A_s$ respectively approach 0, 1 and $\infty$ when $n$ approaches 2.

The numerical results for the $n\leq 2$ scenario are shown in Fig. \ref{fig:n_less_2} (a) $r$ vs. $n_s$, (b)  $r$ vs.  $\log_{10}(\alpha)$, (c) $n_s$ vs. $\log_{10}(\alpha)$, and (d) $\Lambda^4$ vs.  $\log_{10}(\alpha)$, with $n=3/2$ (blue)  $n = 7/4$ (green) and $n = 2$ (red) with $M = 3$ (thick), $M = 4$ (dashed), $M = 5$ (dash-dotted), and $M = 10$ (dotted). The black dots show the predictions of the original natural inflation model. All observables are computed at $N_e = 60$. The amplitude of the power spectrum $A_s$ is fixed to its observed value $A_s \sim 2.1\cdot 10^{-9}$ \cite{Planck2018:inflation}. The gray regions indicate the 95\% (dark-gray) and 68\% (light-gray) confidence levels (CL), respectively, based on the latest combination of Planck, BICEP/Keck, and BAO data \cite{BICEP:2021xfz}.

Panel~(a) of Fig.~\ref{fig:n_less_2} displays the tensor-to-scalar ratio \( r \) plotted against the scalar spectral index \( n_s \). For every choice of $(n, M)$, we scan over \( \alpha \) and fix $\Lambda^4$ in such a way to have $A_s \sim 2.1\cdot 10^{-9}$. The resulting trajectories illustrate how the interplay between the additional \( \alpha R_X^n \) term and the natural inflation potential modifies the prediction for \( r \) and \( n_s \).

Fig.~\ref{fig:n_less_2} (b) shows the behavior of the tensor-to-scalar ratio \(r\) as $\alpha$ varies for three choices of \(n\) and \(M\), the gray-shaded bands represent projections of the 95\% (dark-gray) and 68\% (light-gray) confidence levels based on Planck, BICEP/Keck, and BAO data~\cite{BICEP:2021xfz}.
A key feature is that, for each \(n\), increasing \(\alpha\) generally drives \(r\) towards the allowed regions. 

In panel (c) of Fig.~\ref{fig:n_less_2}, we display the scalar spectral index $n_s$ against $\log_{10}(\alpha)$ for different choices of \(n\) and \(M\), while the shaded regions (dark-gray and light-gray) show the projections of the 95\% and 68\% confidence levels from Planck, BICEP/Keck, and BAO data~\cite{BICEP:2021xfz}. In general, we see that, as \(\alpha\) increases, the $n_s$ value is pushed towards the allowed regions. Different $M$ predict different trajectories for $n_s$.

We notice that for the $n=2$ case in the $\alpha\rightarrow\infty$ the trajectories converge to a single point. This can be easily computed taking the corresponding limit in  \eqref{eq:r:RX2:limit}-\eqref{eq:ns:RX2:limit}, obtaining
\begin{align}
r &\sim 0, \label{eq:r:limit}\\ 
n_s &\sim 1-\frac{3}{2N_e} \label{eq:ns:limit}
\end{align}
where the dependence on $M$ disappears, predicting $r=0, n_s = 0.975$ for $N_e =60$, in agreement with Fig. \ref{fig:n_less_2}. Moreover, the big $\alpha$ limit in eqs. \eqref{eq:r:n:less:2}-\eqref{eq:As:r:n:less:2} for $n<2$ is confirmed as well by the results of Fig. \ref{fig:n_less_2}.

To conclude, we make a brief comment about the possibility of a subPlanckian $M$ in the $n=2$ case. 
It is well known that standard natural inflation suffers of the $\eta$-problem at small $M$'s. This can be easily seen by evaluating the second potential slow-roll parameter (see eq.~\eqref{eq:eta}) for $\phi = 0$:
\begin{eqnarray}
    \eta_U(0) &=&  -\frac{1}{2 M^2 \left(1+16 \alpha  \Lambda ^4\right)} \,.\label{eq:eta:0_exp:full} 
\end{eqnarray}
Therefore, if $\alpha = 0$, we have $|\eta_U(0)| = \frac{1}{2 M^2} > 1$ for $M < 1/\sqrt 2 $, which is in violation of the slow-roll condition $|\eta| \ll 1$. On the other hand, at given $M < 1/\sqrt 2 $ and $\Lambda$, the validity of slow-roll is restored when $\alpha \gg \frac{1-2 M^2}{32 \Lambda ^4 M^2}$. In such a case, then it is possible to obtain results with a behaviour similar to the $M=3$ case shown in Fig. \ref{fig:n_less_2}, but moved to a much lower $r$. For instance, if $\alpha \simeq 6.45 \cdot 10^{13}$ and $M=0.01$, then we obtain for $N_e=60$, $r \simeq 6 \cdot 10^{-8}$, $n_s \simeq 0.964$ and $\Lambda \simeq 0.0044$. For the same $M$ value, the limit in eqs. \eqref{eq:r:limit}-\eqref{eq:ns:limit} can be instead obtained for $\alpha \simeq 2 \cdot 10^{14}.$
\section{Natural inflation for $n > 2$}\label{sec:n>2}
\begin{figure}[t]
    \centering
    \includegraphics[width=0.45\textwidth]{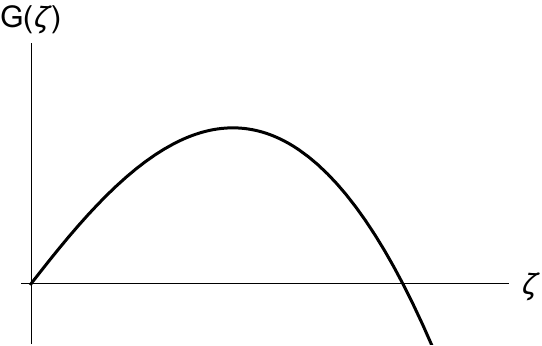}
    \qquad
    \includegraphics[width=0.45\textwidth]{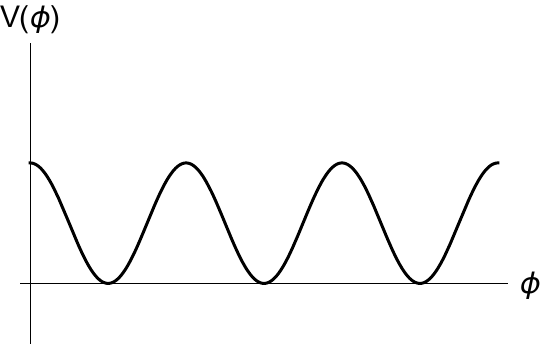}
    \caption{Reference plots of a $G(\zeta)$ (left) generated by a $\FTtwo$ and natural inflation $V(\phi)$ (right). Notice that $G(\zeta)=V(\phi)$ is satisfied easily for any $\phi$ when $\zeta \leq \zeta_0$, provided that the local maximum of $G \geq $ is higher than the maximum of $V$.}
    \label{fig:GvsV}
\end{figure}

The computations proceed in the same way as for the $n<2$ case. However, some fundamental differences appear as well. First of all, the asymptotic values of the CMB observables are not anymore given by \eqref{eq:r:n:less:2}-\eqref{eq:As:r:n:less:2}, because, as explained in the following, the $\alpha\rightarrow\infty$ regime cannot be achieved. We then use the full expressions presented in the appendix (see \eqref{eq:epsilon_full}-\eqref{eq:As_full}).
However, there are first some important considerations to be done, since as it was shown in \cite{Dioguardi:2021fmr}, the case $n \leq 2$ and $n>2$ are conceptually very different.
We see from \eqref{eq:G_n} that the function $G(\zeta)$ is positive only between $\zeta=0$ and $\zeta=\zeta_0=\qty(\alpha(n-2))^{1/1-n}$ (i.e. the solution of $G(\zeta)=0$). Moreover $G(\zeta)$ (see Fig. \ref{fig:GvsV}) also has a maximum in $\zeta_\text{max}$ with
\begin{align}
\zeta_\text{max} &= \qty(\alpha (n-2)n)^\frac{1}{1-n}, \label{eq:zeta:M}
\\
G(\zeta_\text{max}) &= \frac{1}{4} \left(1-\frac{1}{n}\right) (\alpha  (n-2) n)^{\frac{1}{1-n}}=\frac{n-1}{4n}\zeta_\text{max}.\label{eq:G:zeta:M}
\end{align}
\begin{figure}[t]%
    \centering
    {\includegraphics[width=0.7\textwidth]{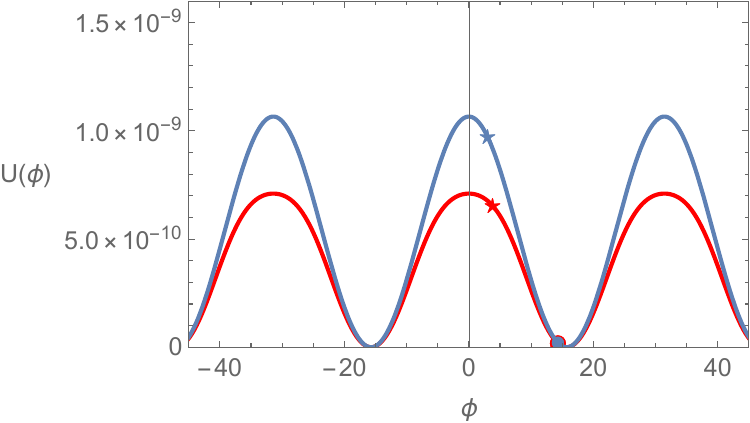}}%
    \caption{Einstein frame potential for the model $F(R_X)=R_X+\alpha R_X^3 $ with $M=5$, and $\alpha=0$ (i.e. the original natural inflation potential) (blue), and for $\alpha_{max}=1.1\cdot 10^{16}$ (red). We also show for reference the field values $\phi_N$ (star) and $\phi_{end}$ (dot). We notice that even for $\alpha=\alpha_{max}$ the potential change is minor. As a consequence, the CMB observables do not change enough, giving $r= 0.02, n_s =0.957$, still outside the experimental region.}
\label{fig:U_potential_3}
\end{figure}
\begin{figure}[t!]%
    \centering
    \subfloat[]{\includegraphics[width=0.47\textwidth]{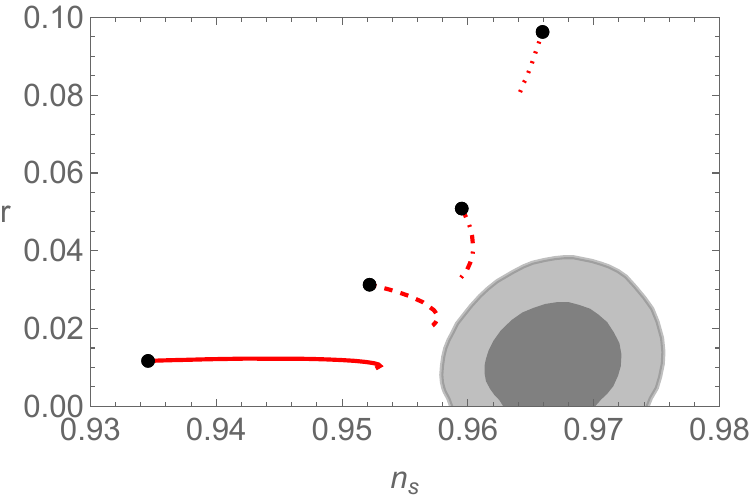}}%
    \qquad
    \subfloat[]{\includegraphics[width=0.45\textwidth]{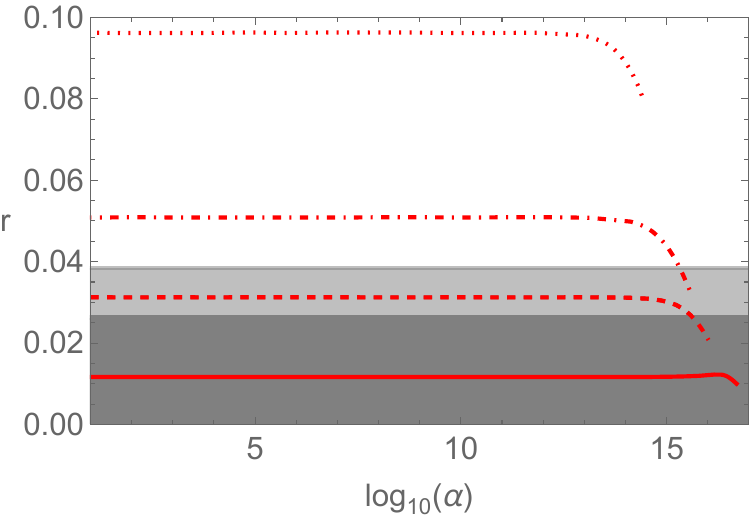}}%
    
    \subfloat[]{\includegraphics[width=0.45\textwidth]{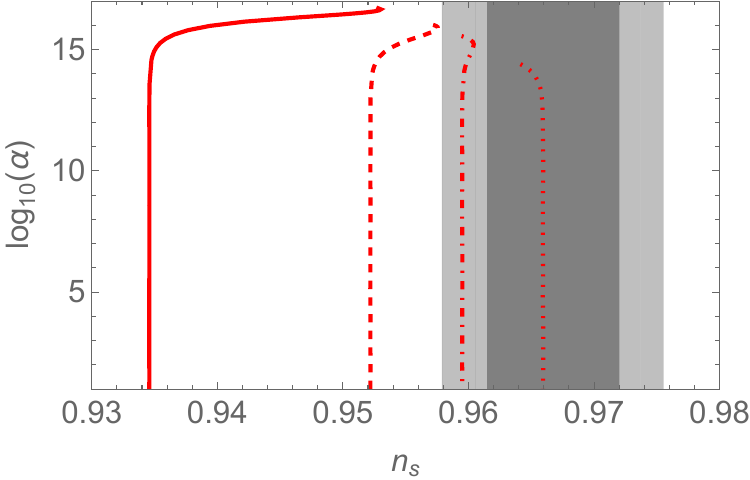}}%
    \qquad
    \subfloat[]{\includegraphics[width=0.455\textwidth]{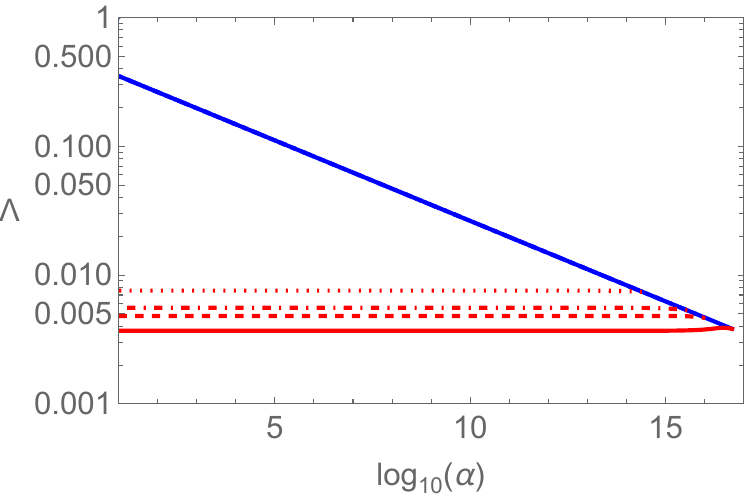}}%

   \caption{(a) \( r \) vs. \( n_s \), (b) \( r \) vs. \( \log_{10}( \alpha) \), (c) \( n_s \) vs. \( \log_{10}(\alpha) \), and (d) \( \Lambda\) vs. \( \log_{10}(\alpha) \) for the model presented in section \ref{sec:n>2}, with \( M = 4 \) (thick), \( M = 5 \) (dashed), \( M = 6 \) (dashed-dotted), and \( M = 10 \) (dotted) and $n=3$. The black dots show the prediction for the original natural inflation potential. The blue line in (d) is $\qty(\frac{G(\zeta_\text{max})}{2})^{1/4}$ as a function of $\alpha$ to make manifest that the model is only defined for $\alpha<\alpha_{max}$. The amplitude of the power spectrum $A_s$ is fixed to its observed value. The gray regions indicate the 95\% (dark-gray) and 68\% (light-gray) confidence levels (CL), respectively, based on the latest combination of Planck, BICEP/Keck, and BAO data \cite{BICEP:2021xfz}.}
    \label{fig:n_bigger_2}
\end{figure}
First of all, it is very important to notice that since we want to preserve the validity of eq. \eqref{eq:EoMzetafull} for all $\phi$ values, we need $G(\zeta_\text{max})\geq2\Lambda^4$, that is we want the maximum of the $G(\zeta)$ function to be higher than the maximum of the Jordan frame potential $V(\phi)$. Unfortunately, this cannot always be true since as $\alpha$ increases, $G(\zeta_\text{max})$ decreases (see \eqref{eq:G:zeta:M}). 

Therefore, at a given $\Lambda$, $\alpha$ must be smaller (or equal) than a $\Lambda$ dependent $\alpha_\text{max}$:
\be
\alpha_\text{max}= \frac{1}{n(n-2)}\qty(\frac{n-1}{4n\Lambda})^{n-1}. \label{eq:alpha:max}
\ee
Therefore, without loss of generality, we start by considering the closest integer number to 2: $n=3$. We show in Fig. \ref{fig:U_potential_3} the Einstein frame potential for a benchmark point, and in Fig. \ref{fig:n_bigger_2} the numerical results for (a) \( r \) vs. \( n_s \), (b) \( r \) vs. \( \log_{10}( \alpha) \), (c) \( n_s \) vs. \( \log_{10}(\alpha) \), and (d) \( \Lambda\) vs. \( \log_{10}(\alpha) \) for the model presented in section \ref{sec:n>2}, with \( M = 4 \) (thick), \( M = 5 \) (dashed), \( M = 6 \) (dashed-dotted), and \( M = 10 \) (dotted) and $n=3$. The black dots show the prediction for the original natural inflation potential. The blue line in (d) is $\qty(\frac{G(\zeta_\text{max})}{2})^{1/4}$ as a function of $\alpha$ that we plot $2\Lambda^4$ to make manifest that the model is only defined for $\alpha\leq\alpha_\text{max}$ (i.e. for $G(\zeta_\text{max}) \geq 2 \Lambda^4$). The amplitude of the power spectrum $A_s$ is fixed to its observed value. The gray regions indicate the 95\% (dark-gray) and 68\% (light-gray) confidence levels (CL), respectively, based on the latest combination of Planck, BICEP/Keck, and BAO data \cite{BICEP:2021xfz}. Since $\alpha$ can't increase to $\alpha\rightarrow \infty$, the predictions do improve with respect to the original natural inflation potential but not enough to reach the allowed region. This feature is actually shared between all the possible $F(R_X)_{>2}$ theories. As can be seen from eq. \eqref{eq:alpha:max}, the closer $n$ to 2, the higher $\alpha_\text{max}$. Therefore we can consider the case in which $n=2+\delta$ with $\delta$ a small effective positive correction induced by a generic $F(R_X)_{>2}$ in the proximity of $\zeta_\text{max}$. In this case \eqref{eq:G_n} is given by:
\be
G(\zeta) = \frac{1}{4}\zeta\qty(1-\alpha \delta \zeta^{1+\delta}) \label{eq:Gdelta}
\ee
while,
\begin{equation}
    \zeta_\text{max} =\qty(\frac{1}{2\alpha\delta(2+\delta)})^{1/1+\delta} \, .
\end{equation}
\newline
In the limit for $\delta\rightarrow 0$ we get:
\be
G(\zeta_\text{max}) \simeq \frac{1}{16\alpha\delta}. \label{eq:Gmaxdelta}
\ee
Hence, in order to counterbalance the high $\alpha$ suppression of the $G(\zeta_\text{max})$ value, we would need an unnatural $\delta \lesssim \frac{1}{\alpha}$. Therefore, unless \emph{ad hoc} constructed cases, the $F(R_X)_{>2}$ framework cannot rescue the natural inflation scenario. Moreover, since eqs. \eqref{eq:Gdelta}-\eqref{eq:Gmaxdelta} do not depend on the specific choice of $V(\phi)$, a similar conclusion can be drawn for any other hilltop-like inflation scenario.

\section{Conclusions}\label{sec:conclusion}
We studied the case of natural inflation in the context of Palatini $F(R,X)$ theories. We considered the $F(R,X)$ dependence to take the simple form $F(R,X) = F(R+X)\equiv F(R_X)$. We then considered two different classes of theories, those with an $R_X$ term diverging faster than quadratic and those which diverge slower. The $n<2$ studied cases have shown to improve the results of standard natural inflation. This especially happens for choice of parameter $3\leq M \leq5$ and for large enough values of $\alpha$, i.e. when the higher order term, $\alpha R_X^n$, becomes relevant during inflation. On the other hand, when $n=2$, the predictions for any $M$ value (also subPlanckian) appear to enter the 2$\sigma$ allowed region for big enough $\alpha$. However, this cannot be achieved for $F(R_X)$ that diverges faster than the quadratic case. If we consider the observed $A_s$, the relation between the higher-order coupling $\alpha$ and the $\Lambda^4$ scale is constrained in such a way that $\alpha$ can never be big enough to contribute to inflation in a sufficient way. In other words, the Einstein frame potential is not substantially modified, and the higher order term represents just an insufficient correction for the prediction of the CMB observables. Summarizing, $F(R_X)_{\leq 2}$ theories appear to restore the viability of natural inflation and the forthcoming CMB experiments, with a precision of $\delta r \sim 10^{-3}$, such as Simons Observatory~\cite{SimonsObservatory:2018koc}, CMB-S4~\cite{Abazajian:2019eic} and LITEBIRD~\cite{LiteBIRD:2020khw}, will be capable test our scenario, specially for the configurations away from the $r \sim 0$ limit.
\acknowledgments
This work was supported by the Estonian Research Council grants PRG1055,  RVTT3, RVTT7 and the CoE program TK202 ``Foundations of the Universe'’.  This article is based upon work from COST Actions COSMIC WISPers CA21106 and
CosmoVerse CA21136, supported by COST (European Cooperation in Science and Technology).

\appendix
\section{More details about slow-roll computations} \label{appendix:SR}

If the Einstein frame potential can be written in terms of the canonical scalar field ($\phi$), by using slow-roll parameters, inflationary predictions can be acquired. The slow-roll parameters with regard to $\phi$ are as follows:
\begin{equation}\label{slowroll1}
\epsilon_U (\phi) =\frac{1}{2}\left( \frac{U(\phi)' }{U(\phi)}\right) ^{2}\,, \quad
\eta_U (\phi) = \frac{U(\phi)^{''}}{U(\phi)} \,,
\end{equation}
here the superscripts $'$ indicate derivatives with respect to the argument, and it is worth emphasizing that we are setting \(M_P\) to unity. In the slow-roll approximation, the inflationary parameters can be defined in the form:
\begin{eqnarray}\label{nsralpha1}
n_s  = 1 - 6 \epsilon_U (\phi_N) + 2 \eta_U (\phi_N)\,,\quad
r = 16 \epsilon_U (\phi_N) \,.
\end{eqnarray} 
where $\phi_N$ is determined by inverting the equation for the number of e-folds which, in the slow-roll approximation, reads
\begin{equation} \label{efold1}
N_e=\int^{\phi_N}_{\phi_{end}}\frac{U(\phi)}{U(\phi)^{'}}d\phi\,. 
\end{equation}
where $\phi_\text{end}$ is the value of the inflaton, where the inflation halts. It can be computed by using the expression, $\epsilon(\phi_\text{end}) = 1$. The number of e-folds is usually taken around 50-60. 
Moreover, the amplitude of curvature perturbations in terms of $\phi$ can be expressed by using the form
\begin{equation} \label{perturb1}
A_s=\frac{1}{24\pi^2}\frac{U(\phi_N)}{\epsilon_U(\phi_N)}
\end{equation}
which should be matched with $A_s\approx 2.1\times10^{-9}$ from the Planck results \cite{Planck2018:inflation} for the pivot scale $k_N = 0.002$ Mpc$^{-1}$. 

The full formulas for the slow-roll parameters and CMB observables for  $V(\phi)$ given by eq. \eqref{eq:V} and $F(R) =R +\alpha R^n$ can be easily obtained by using the method introduced in \cite{Dioguardi:2021fmr} i.e. we use $\zeta$ as the computational variable, we write $U$ in terms of $\zeta$ following eq. \eqref{eq:Uchizeta} and then compute terms like $U(\phi)'$ as $\frac{U(\zeta)'}{\phi(\zeta)'}$, where  $\phi (\zeta)$ is obtained by solving eq. \eqref{eq:EoMzetafull} in terms of $\phi$. After performing all the computations, the slow-roll parameters read:
\begin{align} \label{eq:epsilon_full}
\epsilon_U(\zeta_N) &= \frac{\left(\zeta_N +\alpha  (2-n) \zeta_N ^n\right) \left(8 \Lambda ^4-\zeta_N -\left(\alpha  (2-n) \zeta_N ^n\right)\right)}{2 M^2 \left(\zeta_N +\alpha  n \zeta_N ^n\right)^2}, \\
\eta_U(\zeta_N)=& \frac{\zeta_N ^2 \left(4 \Lambda ^4-\zeta_N\right)+\alpha  \zeta_N ^{n+1} \left(\zeta_N  (n (3 n-4)-2)+4 \Lambda ^4 n (7-5 n)\right)}{M^2 \left(\zeta_N +\alpha  n \zeta_N ^n\right)^2 \left(\zeta_N +\alpha  (2-n) n \zeta_N ^n\right)}+\\ \nonumber
&+\frac{ (2-n) n \alpha ^2\zeta_N ^{2 n} \left(\zeta_N  (4 n-7)+4 \Lambda ^4 (4-3 n)\right)- (2-n)^3 n\alpha ^3 \zeta_N ^{3 n}}{M^2 \left(\zeta_N +\alpha  n \zeta_N ^n\right)^2 \left(\zeta_N +\alpha  (2-n) n \zeta_N ^n\right)}.
\end{align}
Hence, the number of e-folds is given by:
\be
N_e =\int_{\zeta_{end}}^{\zeta_N}\frac{M^2 \left(\zeta +\alpha  n \zeta ^n\right) \left(\zeta +\alpha  (2-n) n \zeta ^n\right)}{\zeta  \left(\zeta +\alpha  (2-n) \zeta ^n\right) \left(-\zeta +8 \Lambda ^4-\left(\alpha  (2-n) \zeta ^n\right)\right)} d\zeta.
\ee
and the CMB observables are:
\begin{align} 
r(\zeta_N) &= 16\epsilon_U(\zeta_N) , \\
n_s(\zeta_N) &=1+2\eta_U(\zeta_N)-6\epsilon_U(\zeta_N), \\
\label{eq:As_full}
A_s(\zeta_N) &= \frac{\zeta_N ^2 M^2 \left(\zeta_N +\alpha  n \zeta_N ^n\right)}{48 \pi ^2 \left(\zeta_N +\alpha  (2-n) \zeta_N ^n\right) \left(-\zeta_N +8 \Lambda ^4-\left(\alpha  (2-n) \zeta_N ^n\right)\right)}.
\end{align}
\bibliographystyle{JHEP}
\bibliography{references}

\end{document}